\documentclass{article}
\begin{document}

\title{Neutrino and it's lepton}
\author{G.Quznetsov\footnote{e-mails: gunn@mail.ru, quznets@yahoo.com}\\
%EndAName
Chelyabinsk State University}
\maketitle

\begin{abstract}
In this paper I cite p.p. 100-117 of book G. Quznetsov, \textit{Probabilistic
Treatment of Gauge Theories}, in series Contemporary Fundamental Physics,
ed. V. Dvoeglazov, Nova Sci. Publ., NY (2007). There I  research a bound
between neutrino and it's lepton.
\end{abstract}

Let $\Im _{e\nu }$ be the unitary space, spanned by the following basis:

\begin{equation}
\mathbf{J}_{e\nu }:=\left\langle 
\begin{array}{c}
\frac{\mathrm{h}}{2\pi \mathrm{c}}\sqrt{\frac{2\pi n_0}{\sinh \left( 2n_0\pi
\right) }}\left( \cosh \left( \frac{\mathrm{h}}{\mathrm{c}}n_0x_4\right)
+\sinh \left( \frac{\mathrm{h}}{\mathrm{c}}n_0x_4\right) \right) \epsilon _1,
\\ 
\frac{\mathrm{h}}{2\pi \mathrm{c}}\sqrt{\frac{2\pi n_0}{\sinh \left( 2n_0\pi
\right) }}\left( \cosh \left( \frac{\mathrm{h}}{\mathrm{c}}n_0x_4\right)
+\sinh \left( \frac{\mathrm{h}}{\mathrm{c}}n_0x_4\right) \right) \epsilon _2,
\\ 
\frac{\mathrm{h}}{2\pi \mathrm{c}}\sqrt{\frac{2\pi n_0}{\sinh \left( 2n_0\pi
\right) }}\left( \cosh \left( \frac{\mathrm{h}}{\mathrm{c}}n_0x_4\right)
-\sinh \left( \frac{\mathrm{h}}{\mathrm{c}}n_0x_4\right) \right) \epsilon _3,
\\ 
\frac{\mathrm{h}}{2\pi \mathrm{c}}\sqrt{\frac{2\pi n_0}{\sinh \left( 2n_0\pi
\right) }}\left( \cosh \left( \frac{\mathrm{h}}{\mathrm{c}}n_0x_4\right)
-\sinh \left( \frac{\mathrm{h}}{\mathrm{c}}n_0x_4\right) \right) \epsilon _4,
\\ 
\frac{\mathrm{h}}{2\pi \mathrm{c}}\exp \left( -\mathrm{i}\frac{\mathrm{h}}{%
\mathrm{c}}\left( n_0x_5\right) \right) \epsilon _1,\frac{\mathrm{h}}{2\pi 
\mathrm{c}}\exp \left( -\mathrm{i}\frac{\mathrm{h}}{\mathrm{c}}\left(
n_0x_5\right) \right) \epsilon _2, \\ 
\frac{\mathrm{h}}{2\pi \mathrm{c}}\exp \left( -\mathrm{i}\frac{\mathrm{h}}{%
\mathrm{c}}\left( n_0x_5\right) \right) \epsilon _3,\frac{\mathrm{h}}{2\pi 
\mathrm{c}}\exp \left( -\mathrm{i}\frac{\mathrm{h}}{\mathrm{c}}\left(
n_0x_5\right) \right) \epsilon _4
\end{array}
\right\rangle \mbox{.}  \label{Jev}
\end{equation}

Let $\Im _e$ be a subspace of the space $\Im _{e\nu }$ such that if $%
\widetilde{\varphi }\in \Im _e$ then $\widetilde{\varphi }$ has got the
following shape:

\[
\widetilde{\varphi }\left( t,\mathbf{x},x_5,x_4\right) =\exp \left( -\mathrm{%
i}\frac{\mathrm{h}}{\mathrm{c}}n_0x_5\right) \sum_{k=1}^4f_k\left( t,\mathbf{%
x},n_0,0\right) \epsilon _k 
\]

That is $\widetilde{\varphi }$ has got the following matrix\index{matrix} in
the basis\index{basis} $\mathbf{J}_{e\nu }$:

\begin{equation}
\widetilde{\varphi }=\left[ 
\begin{array}{c}
0 \\ 
0 \\ 
0 \\ 
0 \\ 
f_1 \\ 
f_2 \\ 
f_3 \\ 
f_4
\end{array}
\right] \mbox{.}  \label{e}
\end{equation}

Let us consider the following Hamiltonian\index{Hamiltonian} on $\Im _e$:

\begin{equation}
\widehat{H}_{0,4}:=\mathrm{c}\left( \sum_{r=1}^3\beta ^{\left[ r\right] }%
\mathrm{i}\partial _r+\gamma ^{\left[ 0\right] }\mathrm{i}\partial _5+\beta
^{\left[ 4\right] }\mathrm{i}\partial _4\right) \mbox{:}  \label{H04}
\end{equation}

\[
\begin{array}{c}
\widehat{H}_{0,4}\widetilde{\varphi }=\mathrm{c}\left( \sum_{r=1}^3\beta
^{\left[ r\right] }\mathrm{i}\partial _r+\gamma ^{\left[ 0\right] }\mathrm{i}%
\partial _5+\beta ^{\left[ 4\right] }\mathrm{i}\partial _4\right) \widetilde{%
\varphi }= \\ 
=\sum_{r=1}^3\beta ^{\left[ r\right] }\mathrm{ci}\partial _r\widetilde{%
\varphi }+ \\ 
+\gamma ^{\left[ 0\right] }\mathrm{ci}\partial _5\exp \left( -\mathrm{i}%
\frac{\mathrm{h}}{\mathrm{c}}n_0x_5\right) \sum_{k=1}^4f_k\left( t,\mathbf{x}%
,n_0,0\right) \epsilon _k+ \\ 
+\beta ^{\left[ 4\right] }\mathrm{ci}\partial _4\exp \left( -\mathrm{i}\frac{%
\mathrm{h}}{\mathrm{c}}n_0x_5\right) \sum_{k=1}^4f_k\left( t,\mathbf{x}%
,n_0,0\right) \epsilon _k= \\ 
=\sum_{r=1}^3\beta ^{\left[ r\right] }\mathrm{ci}\partial _r\widetilde{%
\varphi }+ \\ 
+\gamma ^{\left[ 0\right] }\mathrm{ci}\left( -\mathrm{i}\frac{\mathrm{h}}{%
\mathrm{c}}n_0\right) \exp \left( -\mathrm{i}\frac{\mathrm{h}}{\mathrm{c}}%
n_0x_5\right) \sum_{k=1}^4f_k\left( t,\mathbf{x},n_0,0\right) \epsilon _k+
\\ 
+0= \\ 
=\sum_{r=1}^3\beta ^{\left[ r\right] }\mathrm{ci}\partial _r\widetilde{%
\varphi }+\mathrm{h}n_0\gamma ^{\left[ 0\right] }\exp \left( -\mathrm{i}%
\frac{\mathrm{h}}{\mathrm{c}}n_0x_5\right) \sum_{k=1}^4f_k\left( t,\mathbf{x}%
,n_0,0\right) \epsilon \mbox{.}= \\ 
=\sum_{r=1}^3\beta ^{\left[ r\right] }\mathrm{ci}\partial _r\widetilde{%
\varphi }+\mathrm{h}n_0\gamma ^{\left[ 0\right] }\widetilde{\varphi }\mbox{.}
\end{array}
\]

Hence on this space:

\begin{equation}
\widehat{H}_{0,4}=\widehat{H}_0:=\mathrm{c}\sum_{r=1}^3\beta
^{\left[ r\right] }\mathrm{i}\partial _r+\mathrm{h}n_0\gamma ^{\left[
0\right] }\mbox{.}  \label{H0}
\end{equation}

Let $\Im _{\circ }$ be a subspace of the space $\Im _{e\nu }$ such that if $%
\widetilde{\varphi }_{\circ }\in \Im _{\circ }$ then \\$\widetilde{\varphi }%
_{\circ }=\ell _{\circ }$ $\widetilde{\varphi }$ \cite{Q1} and $\widetilde{%
\varphi }\in \Im _e,$ and if $\widetilde{\varphi }\in \Im _e$ then $\left(
\ell _{\circ }\widetilde{\varphi }\right) \in \Im _{\circ }$. If $\widetilde{%
\varphi }_{\circ }=\ell_{\circ }\widetilde{\varphi }$ then in the
basis\index{basis} $\mathbf{J}_{e\nu }$:

\[
\widetilde{\varphi }_{\circ }=\frac 1{2\sqrt{\left( 1-a^2\right) }}\left[ 
\begin{array}{c}
-\left( -q+ic\right) f_1 \\ 
-\left( -q+ic\right) f_2 \\ 
-\left( -q+ic\right) f_3 \\ 
-\left( -q+ic\right) f_4 \\ 
-\left( -\sqrt{\left( 1-a^2\right) }+b\right) f_1 \\ 
-\left( -\sqrt{\left( 1-a^2\right) }+b\right) f_2 \\ 
-\left( -\sqrt{\left( 1-a^2\right) }+b\right) f_3 \\ 
-\left( -\sqrt{\left( 1-a^2\right) }+b\right) f_4
\end{array}
\right] \mbox{.} 
\]

Let us consider a Hamiltonian\index{Hamiltonian} $\widehat{H}_{0,4}$ mode of
behavior on the space $\Im _{\circ }$:

Hence

\[
\begin{array}{c}
\widehat{H}_{0,4}\widetilde{\varphi }_{\circ }=\mathrm{c}\sum_{r=1}^3\beta
^{\left[ r\right] }\mathrm{i}\partial _r\widetilde{\varphi }_{\circ }+ \\ 
+\gamma ^{\left[ 0\right] }\mathrm{ic}\frac{\left( q-ic\right) }{2\sqrt{%
\left( 1-a^2\right) }}\frac{\mathrm{h}}{2\pi \mathrm{c}}\sqrt{\frac{2\pi n_0%
}{\sinh \left( 2n_0\pi \right) }}\cdot \\ 
\cdot \left( 
\begin{array}{c}
\partial _5\left( 
\begin{array}{c}
f_1\left( \cosh \left( \frac{\mathrm{h}}{\mathrm{c}}n_0x_4\right) +\sinh
\left( \frac{\mathrm{h}}{\mathrm{c}}n_0x_4\right) \right) \epsilon _1+ \\ 
+f_2\left( \cosh \left( \frac{\mathrm{h}}{\mathrm{c}}n_0x_4\right) +\sinh
\left( \frac{\mathrm{h}}{\mathrm{c}}n_0x_4\right) \right) \epsilon _2+ \\ 
+f_3\left( \cosh \left( \frac{\mathrm{h}}{\mathrm{c}}n_0x_4\right) -\sinh
\left( \frac{\mathrm{h}}{\mathrm{c}}n_0x_4\right) \right) \epsilon _3+ \\ 
+f_4\left( \cosh \left( \frac{\mathrm{h}}{\mathrm{c}}n_0x_4\right) -\sinh
\left( \frac{\mathrm{h}}{\mathrm{c}}n_0x_4\right) \right) \epsilon _4+
\end{array}
\right) + \\ 
+\left( \sqrt{\left( 1-a^2\right) }-b\right) \frac{\mathrm{h}}{2\pi \mathrm{c%
}}\partial _5\exp \left( -\mathrm{i}\frac{\mathrm{h}}{\mathrm{c}}\left(
n_0x_5\right) \right) \cdot \\ 
\cdot \left( f_1\epsilon _1+f_2\epsilon _2+f_3\epsilon _3+f_4\epsilon
_4\right)
\end{array}
\right) +
\end{array}
\]

\[
\begin{array}{c}
+\beta ^{\left[ 4\right] }\mathrm{ic}\frac{\left( q-ic\right) }{2\sqrt{%
\left( 1-a^2\right) }}\frac{\mathrm{h}}{2\pi \mathrm{c}}\sqrt{\frac{2\pi n_0%
}{\sinh \left( 2n_0\pi \right) }}\cdot \\ 
\cdot \left( 
\begin{array}{c}
\partial _4\left( 
\begin{array}{c}
f_1\left( \cosh \left( \frac{\mathrm{h}}{\mathrm{c}}n_0x_4\right) +\sinh
\left( \frac{\mathrm{h}}{\mathrm{c}}n_0x_4\right) \right) \epsilon _1+ \\ 
+f_2\left( \cosh \left( \frac{\mathrm{h}}{\mathrm{c}}n_0x_4\right) +\sinh
\left( \frac{\mathrm{h}}{\mathrm{c}}n_0x_4\right) \right) \epsilon _2+ \\ 
+f_3\left( \cosh \left( \frac{\mathrm{h}}{\mathrm{c}}n_0x_4\right) -\sinh
\left( \frac{\mathrm{h}}{\mathrm{c}}n_0x_4\right) \right) \epsilon _3+ \\ 
+f_4\left( \cosh \left( \frac{\mathrm{h}}{\mathrm{c}}n_0x_4\right) -\sinh
\left( \frac{\mathrm{h}}{\mathrm{c}}n_0x_4\right) \right) \epsilon _4+
\end{array}
\right) + \\ 
+\left( \sqrt{\left( 1-a^2\right) }-b\right) \partial _4\frac{\mathrm{h}}{%
2\pi \mathrm{c}}\exp \left( -\mathrm{i}\frac{\mathrm{h}}{\mathrm{c}}\left(
n_0x_5\right) \right) \cdot \\ 
\cdot \left( f_1\epsilon _1+f_2\epsilon _2+f_3\epsilon _3+f_4\epsilon
_4\right)
\end{array}
\right) \mbox{.}
\end{array}
\]

\[
\begin{array}{c}
\widehat{H}_{0,4}\widetilde{\varphi }_{\circ }=\mathrm{c}\sum_{r=1}^3\beta
^{\left[ r\right] }\mathrm{i}\partial _r\widetilde{\varphi }_{\circ }+ \\ 
+\gamma ^{\left[ 0\right] }\mathrm{ic}\frac{\left( q-ic\right) }{2\sqrt{%
\left( 1-a^2\right) }}\frac{\mathrm{h}}{2\pi \mathrm{c}}\sqrt{\frac{2\pi n_0%
}{\sinh \left( 2n_0\pi \right) }}\cdot \\ 
\cdot \left( 
\begin{array}{c}
\partial _5\left( 
\begin{array}{c}
f_1\left( \cosh \left( \frac{\mathrm{h}}{\mathrm{c}}n_0x_4\right) +\sinh
\left( \frac{\mathrm{h}}{\mathrm{c}}n_0x_4\right) \right) \epsilon _1+ \\ 
+f_2\left( \cosh \left( \frac{\mathrm{h}}{\mathrm{c}}n_0x_4\right) +\sinh
\left( \frac{\mathrm{h}}{\mathrm{c}}n_0x_4\right) \right) \epsilon _2+ \\ 
+f_3\left( \cosh \left( \frac{\mathrm{h}}{\mathrm{c}}n_0x_4\right) -\sinh
\left( \frac{\mathrm{h}}{\mathrm{c}}n_0x_4\right) \right) \epsilon _3+ \\ 
+f_4\left( \cosh \left( \frac{\mathrm{h}}{\mathrm{c}}n_0x_4\right) -\sinh
\left( \frac{\mathrm{h}}{\mathrm{c}}n_0x_4\right) \right) \epsilon _4+
\end{array}
\right) + \\ 
+\left( \sqrt{\left( 1-a^2\right) }-b\right) \frac{\mathrm{h}}{2\pi \mathrm{c%
}}\partial _5\exp \left( -\mathrm{i}\frac{\mathrm{h}}{\mathrm{c}}\left(
n_0x_5\right) \right) \cdot \\ 
\cdot \left( f_1\epsilon _1+f_2\epsilon _2+f_3\epsilon _3+f_4\epsilon
_4\right)
\end{array}
\right) +
\end{array}
\]

\[
\begin{array}{c}
+\beta ^{\left[ 4\right] }\mathrm{ic}\frac{\left( q-ic\right) }{2\sqrt{%
\left( 1-a^2\right) }}\frac{\mathrm{h}}{2\pi \mathrm{c}}\sqrt{\frac{2\pi n_0%
}{\sinh \left( 2n_0\pi \right) }}\cdot \\ 
\cdot \left( 
\begin{array}{c}
\partial _4\left( 
\begin{array}{c}
f_1\left( \cosh \left( \frac{\mathrm{h}}{\mathrm{c}}n_0x_4\right) +\sinh
\left( \frac{\mathrm{h}}{\mathrm{c}}n_0x_4\right) \right) \epsilon _1+ \\ 
+f_2\left( \cosh \left( \frac{\mathrm{h}}{\mathrm{c}}n_0x_4\right) +\sinh
\left( \frac{\mathrm{h}}{\mathrm{c}}n_0x_4\right) \right) \epsilon _2+ \\ 
+f_3\left( \cosh \left( \frac{\mathrm{h}}{\mathrm{c}}n_0x_4\right) -\sinh
\left( \frac{\mathrm{h}}{\mathrm{c}}n_0x_4\right) \right) \epsilon _3+ \\ 
+f_4\left( \cosh \left( \frac{\mathrm{h}}{\mathrm{c}}n_0x_4\right) -\sinh
\left( \frac{\mathrm{h}}{\mathrm{c}}n_0x_4\right) \right) \epsilon _4+
\end{array}
\right) + \\ 
+\left( \sqrt{\left( 1-a^2\right) }-b\right) \partial _4\frac{\mathrm{h}}{%
2\pi \mathrm{c}}\exp \left( -\mathrm{i}\frac{\mathrm{h}}{\mathrm{c}}\left(
n_0x_5\right) \right) \cdot \\ 
\cdot \left( f_1\epsilon _1+f_2\epsilon _2+f_3\epsilon _3+f_4\epsilon
_4\right)
\end{array}
\right) \mbox{.}
\end{array}
\]

Therefore

\[
\begin{array}{c}
\widehat{H}_{0,4}\widetilde{\varphi }_{\circ }=\mathrm{c}\sum_{r=1}^3\beta
^{\left[ r\right] }\mathrm{i}\partial _r\widetilde{\varphi }_{\circ }+ \\ 
+\gamma ^{\left[ 0\right] }\mathrm{ic}\frac{\sqrt{1-a^2}-b}{2\sqrt{1-a^2}}%
\cdot \\ 
\cdot \left( 0+\frac{\mathrm{h}}{2\pi \mathrm{c}}\partial _5\exp \left( -%
\mathrm{i}\frac{\mathrm{h}}{\mathrm{c}}\left( n_0x_5\right) \right) \left(
f_1\epsilon _1+f_2\epsilon _2+f_3\epsilon _3+f_4\epsilon _4\right) \right) +
\\ 
+\beta ^{\left[ 4\right] }\mathrm{ic}\frac{q-ic}{2\sqrt{1-a^2}}\frac{\mathrm{%
h}}{2\pi \mathrm{c}}\sqrt{\frac{2\pi n_0}{\sinh \left( 2n_0\pi \right) }}%
\cdot \\ 
\cdot \left( 
\begin{array}{c}
\left( 
\begin{array}{c}
f_1\left( \partial _4\cosh \left( \frac{\mathrm{h}}{\mathrm{c}}n_0x_4\right)
+\partial _4\sinh \left( \frac{\mathrm{h}}{\mathrm{c}}n_0x_4\right) \right)
\epsilon _1+ \\ 
+f_2\left( \partial _4\cosh \left( \frac{\mathrm{h}}{\mathrm{c}}%
n_0x_4\right) +\partial _4\sinh \left( \frac{\mathrm{h}}{\mathrm{c}}%
n_0x_4\right) \right) \epsilon _2+ \\ 
+f_3\left( \partial _4\cosh \left( \frac{\mathrm{h}}{\mathrm{c}}%
n_0x_4\right) -\partial _4\sinh \left( \frac{\mathrm{h}}{\mathrm{c}}%
n_0x_4\right) \right) \epsilon _3+ \\ 
+f_4\left( \partial _4\cosh \left( \frac{\mathrm{h}}{\mathrm{c}}%
n_0x_4\right) -\partial _4\sinh \left( \frac{\mathrm{h}}{\mathrm{c}}%
n_0x_4\right) \right) \epsilon _4+
\end{array}
\right) + \\ 
+0
\end{array}
\right) \mbox{.}
\end{array}
\]

Hence

\[
\begin{array}{c}
\widehat{H}_{0,4}\widetilde{\varphi }_{\circ }=\mathrm{c}\sum_{r=1}^3\beta
^{\left[ r\right] }\mathrm{i}\partial _r\widetilde{\varphi }_{\circ }+ \\ 
+\gamma ^{\left[ 0\right] }\mathrm{ic}\left( -\mathrm{i}\frac{\mathrm{h}}{%
\mathrm{c}}n_0\right) \frac{\sqrt{1-a^2}-b}{2\sqrt{1-a^2}}\frac{\mathrm{h}}{%
2\pi \mathrm{c}}\exp \left( -\mathrm{i}\frac{\mathrm{h}}{\mathrm{c}}%
n_0x_5\right) \cdot \\ 
\cdot \left( f_1\epsilon _1+f_2\epsilon _2+f_3\epsilon _3+f_4\epsilon
_4\right) + \\ 
+\beta ^{\left[ 4\right] }\mathrm{ic}\frac{\mathrm{h}}{\mathrm{c}}n_0\frac{%
q-ic}{2\sqrt{1-a^2}}\frac{\mathrm{h}}{2\pi \mathrm{c}}\sqrt{\frac{2\pi n_0}{%
\sinh \left( 2n_0\pi \right) }}\cdot \\ 
\cdot \left( 
\begin{array}{c}
f_1\left( \sinh \left( \frac{\mathrm{h}}{\mathrm{c}}n_0x_4\right) +\cosh
\left( \frac{\mathrm{h}}{\mathrm{c}}n_0x_4\right) \right) \epsilon _1+ \\ 
+f_2\left( \sinh \left( \frac{\mathrm{h}}{\mathrm{c}}n_0x_4\right) +\cosh
\left( \frac{\mathrm{h}}{\mathrm{c}}n_0x_4\right) \right) \epsilon _2+ \\ 
+f_3\left( \sinh \left( \frac{\mathrm{h}}{\mathrm{c}}n_0x_4\right) -\cosh
\left( \frac{\mathrm{h}}{\mathrm{c}}n_0x_4\right) \right) \epsilon _3+ \\ 
+f_4\left( \sinh \left( \frac{\mathrm{h}}{\mathrm{c}}n_0x_4\right) -\cosh
\left( \frac{\mathrm{h}}{\mathrm{c}}n_0x_4\right) \right) \epsilon _4+
\end{array}
\right) \mbox{.}
\end{array}
\]

Therefore

\[
\begin{array}{c}
\widehat{H}_{0,4}\widetilde{\varphi }_{\circ }=\mathrm{c}\sum_{r=1}^3\beta
^{\left[ r\right] }\mathrm{i}\partial _r\widetilde{\varphi }_{\circ }+ \\ 
+\mathrm{h}n_0\gamma ^{\left[ 0\right] }\frac{\sqrt{1-a^2}-b}{2\sqrt{1-a^2}}%
\frac{\mathrm{h}}{2\pi \mathrm{c}}\exp \left( -\mathrm{i}\frac{\mathrm{h}}{%
\mathrm{c}}n_0x_5\right) \cdot \\ 
\cdot \left( f_1\epsilon _1+f_2\epsilon _2+f_3\epsilon _3+f_4\epsilon
_4\right) + \\ 
+\mathrm{h}n_0\beta ^{\left[ 4\right] }\mathrm{i}\frac{q-ic}{2\sqrt{1-a^2}}%
\frac{\mathrm{h}}{2\pi \mathrm{c}}\sqrt{\frac{2\pi n_0}{\sinh \left( 2n_0\pi
\right) }}\cdot \\ 
\cdot \left( 
\begin{array}{c}
f_1\left( \cosh \left( \frac{\mathrm{h}}{\mathrm{c}}n_0x_4\right) +\sinh
\left( \frac{\mathrm{h}}{\mathrm{c}}n_0x_4\right) \right) \epsilon _1+ \\ 
+f_2\left( \cosh \left( \frac{\mathrm{h}}{\mathrm{c}}n_0x_4\right) +\sinh
\left( \frac{\mathrm{h}}{\mathrm{c}}n_0x_4\right) \right) \epsilon _2- \\ 
-f_3\left( \cosh \left( \frac{\mathrm{h}}{\mathrm{c}}n_0x_4\right) -\sinh
\left( \frac{\mathrm{h}}{\mathrm{c}}n_0x_4\right) \right) \epsilon _3- \\ 
-f_4\left( \cosh \left( \frac{\mathrm{h}}{\mathrm{c}}n_0x_4\right) -\sinh
\left( \frac{\mathrm{h}}{\mathrm{c}}n_0x_4\right) \right) \epsilon _4+
\end{array}
\right) \mbox{.}
\end{array}
\]

Hence in the basis\index{basis} $\mathbf{J}_{e\nu }$:

\[
\begin{array}{c}
\widehat{H}_{0,4}\widetilde{\varphi }_{\circ }=\mathrm{c}\sum_{r=1}^3\beta
^{\left[ r\right] }\mathrm{i}\partial _r\widetilde{\varphi }_{\circ }+%
\mathrm{h}n_0\cdot \\ 
\cdot \left( \gamma ^{\left[ 0\right] }\frac{\sqrt{1-a^2}-b}{2\sqrt{1-a^2}}%
\left[ 
\begin{array}{c}
0 \\ 
0 \\ 
0 \\ 
0 \\ 
f_1 \\ 
f_2 \\ 
f_3 \\ 
f_4
\end{array}
\right] +\beta ^{\left[ 4\right] }\mathrm{i}\frac{q-ic}{2\sqrt{1-a^2}}\left[ 
\begin{array}{c}
f_1 \\ 
f_2 \\ 
-f_3 \\ 
-f_4 \\ 
0 \\ 
0 \\ 
0 \\ 
0
\end{array}
\right] \right) = \\ 
=\mathrm{c}\sum_{r=1}^3\beta ^{\left[ r\right] }\mathrm{i}\partial _r%
\widetilde{\varphi }_{\circ }+\mathrm{h}n_0\cdot \\ 
\cdot \left( \gamma ^{\left[ 0\right] }\frac{\sqrt{1-a^2}-b}{2\sqrt{1-a^2}}%
\left[ 
\begin{array}{c}
0 \\ 
0 \\ 
0 \\ 
0 \\ 
f_1 \\ 
f_2 \\ 
f_3 \\ 
f_4
\end{array}
\right] +\beta ^{\left[ 4\right] }\mathrm{i}\frac{q-ic}{2\sqrt{1-a^2}}\gamma
^{\left[ 5\right] }\left[ 
\begin{array}{c}
f_1 \\ 
f_2 \\ 
f_3 \\ 
f_4 \\ 
0 \\ 
0 \\ 
0 \\ 
0
\end{array}
\right] \right) \mbox{.}
\end{array}
\]

with

\[
\gamma ^{\left[ 5\right] }:=\left[ 
\begin{array}{cccc}
1 & 0 & 0 & 0 \\ 
0 & 1 & 0 & 0 \\ 
0 & 0 & -1 & 0 \\ 
0 & 0 & 0 & -1
\end{array}
\right]\mbox{.} 
\]

Since

\[
\beta ^{\left[ 4\right] }\mathrm{i}\gamma ^{\left[ 5\right] }=\gamma
^{\left[ 0\right] } 
\]

then

\[
\begin{array}{c}
\widehat{H}_{0,4}\widetilde{\varphi }_{\circ }=\mathrm{c}\sum_{r=1}^3\beta
^{\left[ r\right] }\mathrm{i}\partial _r\widetilde{\varphi }_{\circ }+%
\mathrm{h}n_0\cdot \\ 
\cdot \left( \gamma ^{\left[ 0\right] }\frac 1{2\sqrt{1-a^2}}\left[ 
\begin{array}{c}
0 \\ 
0 \\ 
0 \\ 
0 \\ 
\left( \sqrt{1-a^2}-b\right) f_1 \\ 
\left( \sqrt{1-a^2}-b\right) f_2 \\ 
\left( \sqrt{1-a^2}-b\right) f_3 \\ 
\left( \sqrt{1-a^2}-b\right) f_4
\end{array}
\right] +\gamma ^{\left[ 0\right] }\frac 1{2\sqrt{1-a^2}}\left[ 
\begin{array}{c}
\left( q-ic\right) f_1 \\ 
\left( q-ic\right) f_2 \\ 
\left( q-ic\right) f_3 \\ 
\left( q-ic\right) f_4 \\ 
0 \\ 
0 \\ 
0 \\ 
0
\end{array}
\right] \right) \mbox{.}
\end{array}
\]

Therefore

\[
\widehat{H}_{0,4}\widetilde{\varphi }_{\circ }=\mathrm{c}\sum_{r=1}^3\beta
^{\left[ r\right] }\mathrm{i}\partial _r\widetilde{\varphi }_{\circ }+%
\mathrm{h}n_0\gamma ^{\left[ 0\right] }\frac 1{2\sqrt{1-a^2}}\left( 
\begin{array}{c}
-\left( -q+ic\right) f_1, \\ 
-\left( -q+ic\right) f_2, \\ 
-\left( -q+ic\right) f_3, \\ 
-\left( -q+ic\right) f_4, \\ 
-\left( -\sqrt{1-a^2}+b\right) f_1, \\ 
-\left( -\sqrt{1-a^2}+b\right) f_2, \\ 
-\left( -\sqrt{1-a^2}+b\right) f_3, \\ 
-\left( -\sqrt{1-a^2}+b\right) f_4
\end{array}
\right) \mbox{.} 
\]

Hence

\[
\widehat{H}_{0,4}\widetilde{\varphi }_{\circ }=\mathrm{c}\sum_{r=1}^3\beta
^{\left[ r\right] }\mathrm{i}\partial _r\widetilde{\varphi }_{\circ }+%
\mathrm{h}n_0\gamma ^{\left[ 0\right] }\widetilde{\varphi }_{\circ }\mbox{.} 
\]

Thus in the space $\Im _e$:

\[
\widehat{H}_{0,4}=\widehat{H}_0=\mathrm{c}\sum_{r=1}^3\beta ^{\left[
r\right] }\mathrm{i}\partial _r+\mathrm{h}n_0\gamma ^{\left[ 0\right] }%
\mbox{,} 
\]

too.

Let $\Im _{*}$ be a subspace of the space $\Im _{e\nu }$ such that if $%
\widetilde{\varphi }_{*}\in \Im _{*}$ then \\$\widetilde{\varphi }_{*}=\ell
_{*}$ $\widetilde{\varphi }$ \cite{Q1} and $\widetilde{\varphi }\in \Im _e,$
and if $\widetilde{\varphi }\in \Im _e$ then $\left( \ell _{*}\widetilde{%
\varphi }\right) \in \Im _{*}$. If $\widetilde{\varphi }_{*}=\ell _{*}%
\widetilde{\varphi }$ (\ref{e}) then in the basis\index{basis} $\mathbf{J}%
_{e\nu }$:

\[
\widetilde{\varphi }_{*}=\frac 1{2\sqrt{\left( 1-a^2\right) }}\left[ 
\begin{array}{c}
\left( -q+ic\right) f_1 \\ 
\left( -q+ic\right) f_2 \\ 
\left( -q+ic\right) f_3 \\ 
\left( -q+ic\right) f_4 \\ 
\left( b+\sqrt{1-a^2}\right) f_1 \\ 
\left( b+\sqrt{1-a^2}\right) f_2 \\ 
\left( b+\sqrt{1-a^2}\right) f_3 \\ 
\left( b+\sqrt{1-a^2}\right) f_4
\end{array}
\right] \mbox{.} 
\]

Similarly to $\widetilde{\varphi }_{\circ }$ you can calculate that

\[
\widehat{H}_{0,4}\widetilde{\varphi }_{*}=\widehat{H}_0\widetilde{\varphi }%
_{*}=\mathrm{c}\sum_{r=1}^3\beta ^{\left[ r\right] }\mathrm{i}\partial _r%
\widetilde{\varphi }_{*}+\mathrm{h}n_0\gamma ^{\left[ 0\right] }\widetilde{%
\varphi }_{*}.\mbox{,} 
\]

too.

Let

\[
e_{1L}\left( \mathbf{k}\right) :=\left[ 
\begin{array}{c}
\omega \left( \mathbf{k}\right) +n_0+k_3 \\ 
k_1+\mathrm{i}k_2
\end{array}
\right] \mbox{, }e_{1R}\left( \mathbf{k}\right) :=\left[ 
\begin{array}{c}
\omega \left( \mathbf{k}\right) +n_0-k_3 \\ 
-k_1-\mathrm{i}k_2
\end{array}
\right] \mbox{,} 
\]

\[
e_{2L}\left( \mathbf{k}\right) :=\left[ 
\begin{array}{c}
k_1-\mathrm{i}k_2 \\ 
\omega \left( \mathbf{k}\right) +n_0-k_3
\end{array}
\right] \mbox{, }e_{2R}\left( \mathbf{k}\right) :=\left[ 
\begin{array}{c}
-k_1+\mathrm{i}k_2 \\ 
\omega \left( \mathbf{k}\right) +n_0+k_3
\end{array}
\right] \mbox{,} 
\]

\begin{eqnarray*}
&&e_{3L}\left( \mathbf{k}\right) :=-e_{1R}\left( \mathbf{k}%
\right) \mbox{, }e_{3R}\left( \mathbf{k}\right) :=%
e_{1L}\left( \mathbf{k}\right) \mbox{, } \\
&&e_{4L}\left( \mathbf{k}\right) :=-e_{2R}\left( \mathbf{k}%
\right) \mbox{, }e_{4R}\left( \mathbf{k}\right) :=%
e_{2L}\left( \mathbf{k}\right) \mbox{.}
\end{eqnarray*}

with

\[
\omega \left( \mathbf{k}\right) :=\sqrt{%
n_0^2+k_1^2+k_2^2+k_3^2} 
\]

( $n_0$, $k_1$, $k_2$,$k_3$ are real numbers).

In this case \cite{Q2}:

\[
e_s\left( \mathbf{k}\right) =\frac 1{2\sqrt{\omega \left( \mathbf{k}\right)
\left( \omega \left( \mathbf{k}\right) +n_0\right) }}\left[ 
\begin{array}{c}
e_{sL}\left( \mathbf{k}\right) \\ 
e_{sR}\left( \mathbf{k}\right)
\end{array}
\right] \mbox{.} 
\]

Let

\[
\underline{e}_s\left( \mathbf{k}\right) :=\left[ 
\begin{array}{c}
\overrightarrow{0}_4 \\ 
e_s\left( \mathbf{k}\right)
\end{array}
\right] 
\]

here $s\in \left\{ 1,2,3,4\right\} $.

And let \cite{Q3}:

\begin{eqnarray*}
&&\underline{e}_{\circ s}\left( \mathbf{k}\right) :=\ell
_{\circ }\underline{e}_s\left( \mathbf{k}\right) \\
&=&\frac 1{\sqrt{2\left( \sqrt{1-a^2}-b\right) \sqrt{1-a^2}}}\left[ 
\begin{array}{c}
\left( q-\mathrm{i}c\right) e_s\left( \mathbf{k}\right) \\ 
\left( \sqrt{1-a^2}-b\right) e_s\left( \mathbf{k}\right)
\end{array}
\right] \mbox{,}
\end{eqnarray*}

\begin{eqnarray*}
&&\underline{e}_{*s}\left( \mathbf{k}\right) :=\ell _{*}%
\underline{e}_s\left( \mathbf{k}\right) \\
&=&\frac 1{\sqrt{2\left( \sqrt{1-a^2}+b\right) \sqrt{1-a^2}}}\left[ 
\begin{array}{c}
\left( -q+\mathrm{i}c\right) e_s\left( \mathbf{k}\right) \\ 
\left( b+\sqrt{1-a^2}\right) e_s\left( \mathbf{k}\right)
\end{array}
\right] \mbox{.}
\end{eqnarray*}

Denote

\[
\widehat{H}_0\left( \mathbf{k}\right) :=\sum_{r=1}^3\beta
^{\left[ r\right] }k_r=\left[ 
\begin{array}{cccc}
k_3 & k_1-\mathrm{i}k_2 & n_0 & 0 \\ 
k_1+\mathrm{i}k_2 & -k_3 & 0 & n_0 \\ 
n_0 & 0 & -k_3 & -k_1+\mathrm{i}k_2 \\ 
0 & n_0 & -k_1-\mathrm{i}k_2 & k_3
\end{array}
\right] \mbox{.} 
\]

In that case

\begin{eqnarray*}
&&\widehat{H}_0\underline{e}_{\circ 1}\left( \mathbf{k}\right) \left( \frac{%
\mathrm{h}}{2\pi \mathrm{c}}\right) ^{\frac 32}\exp \left( \mathrm{i}\frac{%
\mathrm{h}}{\mathrm{c}}\right)= \\
&=&\mathrm{h}\widehat{H}_0\left( \mathbf{k}\right) \underline{e}_{\circ
1}\left( \mathbf{k}\right) \left( \frac{\mathrm{h}}{2\pi \mathrm{c}}\right)
^{\frac 32}\exp \left( \mathrm{i}\frac{\mathrm{h}}{\mathrm{c}}\right) \\
&=&\mathrm{h}\omega \left( \mathbf{k}\right) \underline{e}_{\circ 1}\left( 
\mathbf{k}\right) \left( \frac{\mathrm{h}}{2\pi \mathrm{c}}\right) ^{\frac
32}\exp \left( \mathrm{i}\frac{\mathrm{h}}{\mathrm{c}}\right) \mbox{.}
\end{eqnarray*}

Therefore $\underline{e}_{\circ 1}\left( \mathbf{k}\right) \left( \frac{%
\mathrm{h}}{2\pi \mathrm{c}}\right) ^{\frac 32}\exp \left( \mathrm{i}\frac{%
\mathrm{h}}{\mathrm{c}}\right) $ is an eigenvector\index{eigenvector} of $%
\widehat{H}_0$ with the eigenvalue\index{eigenvalue} $\mathrm{h}\omega
\left( \mathbf{k}\right) $. Similarly you can calculate that \\$\underline{e}%
_{\circ 2}\left( \mathbf{k}\right) \left( \frac{\mathrm{h}}{2\pi \mathrm{c}}%
\right) ^{\frac 32}\exp \left( \mathrm{i}\frac{\mathrm{h}}{\mathrm{c}}%
\right) $, $\underline{e}_{*1}\left( \mathbf{k}\right) \left( \frac{\mathrm{h%
}}{2\pi \mathrm{c}}\right) ^{\frac 32}\exp \left( \mathrm{i}\frac{\mathrm{h}%
}{\mathrm{c}}\right) $, $\underline{e}_{*2}\left( \mathbf{k}\right) \left( 
\frac{\mathrm{h}}{2\pi \mathrm{c}}\right) ^{\frac 32}\exp \left( \mathrm{i}%
\frac{\mathrm{h}}{\mathrm{c}}\right) $, \\are eigenvectors\index{eigenvector}
of $\widehat{H}_0$ with the same eigenvalue, \index{eigenvalue}and \\$%
\underline{e}_{\circ 3}\left( \mathbf{k}\right) \left( \frac{\mathrm{h}}{%
2\pi \mathrm{c}}\right) ^{\frac 32}\exp \left( \mathrm{i}\frac{\mathrm{h}}{%
\mathrm{c}}\right) $, $\underline{e}_{\circ 4}\left( \mathbf{k}\right)
\left( \frac{\mathrm{h}}{2\pi \mathrm{c}}\right) ^{\frac 32}\exp \left( 
\mathrm{i}\frac{\mathrm{h}}{\mathrm{c}}\right) $,\\$\underline{e}_{*3}\left( 
\mathbf{k}\right) \left( \frac{\mathrm{h}}{2\pi \mathrm{c}}\right) ^{\frac
32}\exp \left( \mathrm{i}\frac{\mathrm{h}}{\mathrm{c}}\right) $, $\underline{%
e}_{*4}\left( \mathbf{k}\right) \left( \frac{\mathrm{h}}{2\pi \mathrm{c}}%
\right) ^{\frac 32}\exp \left( \mathrm{i}\frac{\mathrm{h}}{\mathrm{c}}%
\right) $ \\are an eigenvectors\index{eigenvector} of $\widehat{H}_0$ with
the eigenvalue\index{eigenvalue} $\left( -\mathrm{h}\omega \left( \mathbf{k}%
\right) \right) $.

Vectors $\underline{e}_{\circ s}\left( \mathbf{k}\right) \left( \frac{%
\mathrm{h}}{2\pi \mathrm{c}}\right) ^{\frac 32}\exp \left( \mathrm{i}\frac{%
\mathrm{h}}{\mathrm{c}}\right) $, $\underline{e}_{*s}\left( \mathbf{k}%
\right) \left( \frac{\mathrm{h}}{2\pi \mathrm{c}}\right) ^{\frac 32}\exp
\left( \mathrm{i}\frac{\mathrm{h}}{\mathrm{c}}\right) $ with $s\in \left\{
1,2,3,4\right\} $ form an orthonormalized\index{normalized} basis%
\index{basis} in the space $\Im _{e\nu }$ (\ref{Jev}) and

\begin{equation}
\sum_{s=1}^4\left( \underline{e}_{*s,r}^{*}\left( \mathbf{k}\right) 
\underline{e}_{*s,r^{\prime }}\left( \mathbf{k}\right) +\underline{e}_{\circ
s,r}^{*}\left( \mathbf{k}\right) \underline{e}_{\circ s,r^{\prime }}\left( 
\mathbf{k}\right) \right) =\delta _{r,r^{\prime }}  \label{rr}
\end{equation}

for $r,r^{\prime}\in \left\{ 1,2,3,4,5,6,7,8\right\} $.

Let \cite{Q3}

\begin{eqnarray*}
&&\underline{e}_{*s}^{\prime }\left( \mathbf{k}\right) :=%
U^{\left( -\right) }\underline{e}_{*s}\left( \mathbf{k}\right) \\
&=&\frac 1{\sqrt{2\left( \sqrt{1-a^2}+b\right) \sqrt{1-a^2}}}\frac 1{2\sqrt{%
\omega \left( \mathbf{k}\right) \left( \omega \left( \mathbf{k}\right)
+n_0\right) }} \\
&&\cdot \left[ 
\begin{array}{c}
\left( a-\mathrm{i}\sqrt{1-a^2}\right) \left( -q+\mathrm{i}c\right)
e_{sL}\left( \mathbf{k}\right) \\ 
\left( -q+\mathrm{i}c\right) e_{sR}\left( \mathbf{k}\right) \\ 
\left( a-\mathrm{i}\sqrt{1-a^2}\right) \left( \sqrt{1-a^2}+b\right)
e_{sL}\left( \mathbf{k}\right) \\ 
\left( \sqrt{\left( 1-a^2\right) }+b\right) e_{sR}\left( \mathbf{k}\right)
\end{array}
\right]
\end{eqnarray*}

and

\begin{eqnarray*}
&&\underline{e}_{\circ s}^{\prime }\left( \mathbf{k}\right) :=%
U^{\left( -\right) }\underline{e}_{\circ s}\left( \mathbf{k}\right) \\
&=&\frac 1{\sqrt{2\left( \sqrt{1-a^2}-b\right) \sqrt{1-a^2}}}\frac 1{2\sqrt{%
\omega \left( \mathbf{k}\right) \left( \omega \left( \mathbf{k}\right)
+n_0\right) }} \\
&&\cdot \left[ 
\begin{array}{c}
\left( a+\mathrm{i}\sqrt{1-a^2}\right) \left( q-\mathrm{i}c\right)
e_{sL}\left( \mathbf{k}\right) \\ 
\left( q-\mathrm{i}c\right) e_{sR}\left( \mathbf{k}\right) \\ 
\left( a+\mathrm{i}\sqrt{1-a^2}\right) \left( \sqrt{1-a^2}-b\right)
e_{sL}\left( \mathbf{k}\right) \\ 
\left( \sqrt{\left( 1-a^2\right) }-b\right) e_{sR}\left( \mathbf{k}\right)
\end{array}
\right] \mbox{.}
\end{eqnarray*}

For these vectors:

\[
\sum_{r=1}^4\left( \underline{e}_{*r,j}^{\prime *}\left( \mathbf{k}\right) 
\underline{e}_{*r,j^{\prime }}^{\prime }\left( \mathbf{k}\right) +\underline{%
e}_{\circ r,j}^{\prime *}\left( \mathbf{k}\right) \underline{e}_{\circ
r,j^{\prime }}^{\prime }\left( \mathbf{k}\right) \right) =\delta
_{j,j^{\prime }} 
\]

and since $U^{\left( -\right) \dagger }U^{\left( -\right) }=1_8$ then $%
\underline{e}_{\circ s}^{\prime }\left( \mathbf{k}\right) $ and $\underline{e%
}_{*s}^{\prime }\left( \mathbf{k}\right) $ form an orthonormalized%
\index{normalized} basis\index{basis} in the space $\Im _{e\nu }$, too.

Let

\begin{equation}
\underline{e}_r^{\prime }\left( \mathbf{k}\right) :=U^{\left(
-\right) }\underline{e}_r\left( \mathbf{k}\right) =\frac 1{2\sqrt{\omega
\left( \mathbf{k}\right) \left( \omega \left( \mathbf{k}\right) +n_0\right) }%
}\left[ 
\begin{array}{c}
\left( c+\mathrm{i}q\right) e_{rL}\left( \mathbf{k}\right) \\ 
\overrightarrow{0}_2 \\ 
\left( a-\mathrm{i}b\right) e_{rL}\left( \mathbf{k}\right) \\ 
e_{rR}\left( \mathbf{k}\right)
\end{array}
\right] \mbox{.}  \label{ntr2}
\end{equation}

In that case:

\[
\underline{e}_r^{\prime }\left( \mathbf{k}\right) =\frac 1{\sqrt{2}}\left( 
\sqrt{1-\frac b{\sqrt{1-a^2}}}\underline{e}_{\circ r}^{\prime }\left( 
\mathbf{k}\right) +\sqrt{1+\frac b{\sqrt{1-a^2}}}\underline{e}_{*r}^{\prime
}\left( \mathbf{k}\right) \right) \mbox{.} 
\]

Let for $j,j^{\prime }\in \left\{ 1,2,3,4,5,6,7,8\right\} $ \cite{Q4}:

\[
\begin{array}{c}
\left\{ \psi _{j^{\prime }}^{\dagger }\left( \mathbf{y}\right) ,\psi
_j\left( \mathbf{x}\right) \right\} =\delta \left( \mathbf{{{y}-{x}}}\right)
\delta _{j^{\prime },j}\widehat{1}\mbox{,} \\ 
\left\{ \psi _{j^{\prime }}^{\dagger }\left( \mathbf{y}\right) ,\psi
_j^{\dagger }\left( \mathbf{x}\right) \right\} =\widehat{0}=\left\{ \psi
_{j^{\prime }}\left( \mathbf{y}\right) ,\psi _j\left( \mathbf{x}\right)
\right\}
\end{array}
\]

and let

\begin{eqnarray*}
&&\ b_{\circ r,\mathbf{k}}:=\left( \frac{\mathrm{h}}{2\pi 
\mathrm{c}}\right) ^3\int_{\left( \mathbf{\Omega }\right) }d\mathbf{x}\cdot
e^{\mathrm{i}\frac{\mathrm{h}}{\mathrm{c}}\mathbf{kx}}\sum_{j^{\prime }=1}^8%
\underline{e}_{\circ r,j^{\prime }}^{*}\left( \mathbf{k}\right) \psi
_{j^{\prime }}\left( \mathbf{x}\right) \mathbf{\mbox{,}} \\
&&\ b_{*r,\mathbf{k}}:=\left( \frac{\mathrm{h}}{2\pi \mathrm{c%
}}\right) ^3\int_{\left( \mathbf{\Omega }\right) }d\mathbf{x}\cdot e^{%
\mathrm{i}\frac{\mathrm{h}}{\mathrm{c}}\mathbf{kx}}\sum_{j^{\prime }=1}^8%
\underline{e}_{*r,j^{\prime }}^{*}\left( \mathbf{k}\right) \psi _{j^{\prime
}}\left( \mathbf{x}\right) \mathbf{\mbox{.}}
\end{eqnarray*}

In that case:

\begin{eqnarray*}
&&\sum_{\mathbf{k}}e^{-\mathrm{i}\frac{\mathrm{h}}{\mathrm{c}}\mathbf{kx}%
}\left( \sum_{r=1}^4\underline{e}_{\circ r,j}\left( \mathbf{k}\right)
b_{\circ r,\mathbf{k}}+\sum_{r=1}^4\underline{e}_{*r,j}\left( \mathbf{k}%
\right) b_{*r,\mathbf{k}}\right) \\
&=&\sum_{\mathbf{k}}e^{-\mathrm{i}\frac{\mathrm{h}}{\mathrm{c}}\mathbf{kx}%
}\left( 
\begin{array}{c}
\sum_{r=1}^4\underline{e}_{\circ r,j}\left( \mathbf{k}\right) \left( \frac{%
\mathrm{h}}{2\pi \mathrm{c}}\right) ^3\cdot \\ 
\cdot \int_{\left( \mathbf{\Omega }\right) }d\mathbf{x}^{\prime }\cdot e^{%
\mathrm{i}\frac{\mathrm{h}}{\mathrm{c}}\mathbf{kx}^{\prime }}\sum_{j^{\prime
}=1}^8\underline{e}_{\circ r,j^{\prime }}^{*}\left( \mathbf{k}\right) \psi
_{j^{\prime }}\left( \mathbf{x}^{\prime }\right) \\ 
+\sum_{r=1}^4\underline{e}_{*r,j}\left( \mathbf{k}\right) \left( \frac{%
\mathrm{h}}{2\pi \mathrm{c}}\right) ^3\cdot \\ 
\cdot \int_{\left( \mathbf{\Omega }\right) }d\mathbf{x}^{\prime }\cdot e^{%
\mathrm{i}\frac{\mathrm{h}}{\mathrm{c}}\mathbf{kx}^{\prime }}\sum_{j^{\prime
}=1}^8\underline{e}_{*r,j^{\prime }}^{*}\left( \mathbf{k}\right) \psi
_{j^{\prime }}\left( \mathbf{x}^{\prime }\right)
\end{array}
\right) \\
&=&\left( \frac{\mathrm{h}}{2\pi \mathrm{c}}\right) ^3\sum_{\mathbf{k}%
}\int_{\left( \mathbf{\Omega }\right) }d\mathbf{x}^{\prime }\cdot e^{\mathrm{%
i}\frac{\mathrm{h}}{\mathrm{c}}\mathbf{kx}^{\prime }}e^{-\mathrm{i}\frac{%
\mathrm{h}}{\mathrm{c}}\mathbf{kx}}\cdot \\
&&\cdot \sum_{r=1}^4\sum_{j^{\prime }=1}^8\left( \underline{e}_{\circ
r,j}\left( \mathbf{k}\right) \underline{e}_{\circ r,j^{\prime }}^{*}\left( 
\mathbf{k}\right) +\underline{e}_{*r,j}\left( \mathbf{k}\right) \underline{e}%
_{*r,j^{\prime }}^{*}\left( \mathbf{k}\right) \right) \psi _{j^{\prime
}}\left( \mathbf{x}^{\prime }\right) \mbox{.}
\end{eqnarray*}

In accordance with (\ref{rr}):

\begin{eqnarray*}
&&\ \sum_{\mathbf{k}}e^{-\mathrm{i}\frac{\mathrm{h}}{\mathrm{c}}\mathbf{kx}%
}\left( \sum_{r=1}^4\underline{e}_{\circ r,j}\left( \mathbf{k}\right)
b_{\circ r,\mathbf{k}}+\sum_{r=1}^4\underline{e}_{*r,j}\left( \mathbf{k}%
\right) b_{*r,\mathbf{k}}\right) \\
&=&\left( \frac{\mathrm{h}}{2\pi \mathrm{c}}\right) ^3\int_{\left( \mathbf{%
\Omega }\right) }d\mathbf{x}^{\prime }\cdot \sum_{\mathbf{k}}e^{-\mathrm{i}%
\frac{\mathrm{h}}{\mathrm{c}}\mathbf{k}\left( \mathbf{x-x}^{\prime }\right)
}\sum_{j^{\prime }=1}^8\delta _{j,j^{\prime }}\psi _{j^{\prime }}\left( 
\mathbf{x}^{\prime }\right) \mbox{.}
\end{eqnarray*}

Hence since

\[
\sum_{\mathbf{k}}e^{\mathrm{i}\frac{\mathrm{h}}{\mathrm{c}}\mathbf{k}\left( 
\mathbf{x}^{\prime }-\mathbf{x}\right) }=\left( \frac h{2\pi \mathrm{c}%
}\right) ^3\delta \left( \mathbf{x}^{\prime }-\mathbf{x}\right) 
\]

and according properties of $\delta $:

\begin{eqnarray*}
&&\ \ \sum_{\mathbf{k}}e^{-\mathrm{i}\frac{\mathrm{h}}{\mathrm{c}}\mathbf{kx}%
}\left( \sum_{r=1}^4\underline{e}_{\circ r,j}\left( \mathbf{k}\right)
b_{\circ r,\mathbf{k}}+\sum_{r=1}^4\underline{e}_{*r,j}\left( \mathbf{k}%
\right) b_{*r,\mathbf{k}}\right) \\
\ &=&\left( \frac{\mathrm{h}}{2\pi \mathrm{c}}\right) ^3\int_{\left( \mathbf{%
\Omega }\right) }d\mathbf{x}^{\prime }\cdot \left( \frac h{2\pi \mathrm{c}%
}\right) ^3\delta \left( \mathbf{x}^{\prime }-\mathbf{x}\right) \psi
_j\left( \mathbf{x}^{\prime }\right) \\
&=&\int_{\left( \mathbf{\Omega }\right) }d\mathbf{x}^{\prime }\cdot \delta
\left( \mathbf{x}^{\prime }-\mathbf{x}\right) \psi _j\left( \mathbf{x}%
^{\prime }\right) =\psi _j\left( \mathbf{x}\right) \mbox{.}
\end{eqnarray*}

Thus:

\begin{equation}
\frame{$\sum_{\mathbf{k}}e^{-\mathrm{i}\frac{\mathrm{h}}{\mathrm{c}}\mathbf{%
kx}}\left( \sum_{r=1}^4\underline{e}_{\circ r,j}\left( \mathbf{k}\right)
b_{\circ r,\mathbf{k}}+\sum_{r=1}^4\underline{e}_{*r,j}\left( \mathbf{k}%
\right) b_{*r,\mathbf{k}}\right) =\psi _j\left( \mathbf{x}\right) $\mbox{.}}
\label{ntr1}
\end{equation}

Let

\begin{equation}
\begin{array}{c}
\psi \left( \mathbf{x}\right) :=\frac{\mathrm{h}}{2\pi 
\mathrm{c}}\cdot \\ 
\cdot \left( 
\begin{array}{c}
\sqrt{\frac{2\pi n_0}{\sinh \left( 2n_0\pi \right) }}\left( 
\begin{array}{c}
\left( 
\begin{array}{c}
\cosh \left( \frac{\mathrm{h}}{\mathrm{c}}n_0x_4\right) + \\ 
+\sinh \left( \frac{\mathrm{h}}{\mathrm{c}}n_0x_4\right)
\end{array}
\right) \sum_{r=1}^2\psi _r\left( \mathbf{x}\right) \epsilon _r+ \\ 
+\left( 
\begin{array}{c}
\cosh \left( \frac{\mathrm{h}}{\mathrm{c}}n_0x_4\right) - \\ 
-\sinh \left( \frac{\mathrm{h}}{\mathrm{c}}n_0x_4\right)
\end{array}
\right) \sum_{r=3}^4\psi _r\left( \mathbf{x}\right) \epsilon _r
\end{array}
\right) + \\ 
+\exp \left( -\mathrm{i}\frac{\mathrm{h}}{\mathrm{c}}\left( n_0x_4\right)
\right) \sum_{r=1}^4\psi _{r+4}\left( \mathbf{x}\right) \epsilon _r
\end{array}
\right) \mbox{.}
\end{array}
\label{ntr4}
\end{equation}

That is in the basis\index{basis} $\mathbf{J}_{e\nu}$ (\ref{Jev}):

\[
\psi \left( \mathbf{x}\right) =\left[ 
\begin{array}{c}
\psi _1\left( \mathbf{x}\right) \\ 
\psi _2\left( \mathbf{x}\right) \\ 
\psi _3\left( \mathbf{x}\right) \\ 
\psi _4\left( \mathbf{x}\right) \\ 
\psi _5\left( \mathbf{x}\right) \\ 
\psi _6\left( \mathbf{x}\right) \\ 
\psi _7\left( \mathbf{x}\right) \\ 
\psi _8\left( \mathbf{x}\right)
\end{array}
\right] \mbox{.} 
\]

That is in this basis\index{basis}:

\begin{eqnarray*}
&&\ b_{\circ r,\mathbf{k}}:=\left( \frac{\mathrm{h}}{2\pi 
\mathrm{c}}\right) ^3\int_{\left( \mathbf{\Omega }\right) }d\mathbf{x}\cdot
e^{\mathrm{i}\frac{\mathrm{h}}{\mathrm{c}}\mathbf{kx}}\underline{e}_{\circ
r,j}^{\dagger }\left( \mathbf{k}\right) \psi \left( \mathbf{x}\right) 
\mathbf{\mbox{,}} \\
&&\ b_{*r,\mathbf{k}}:=\left( \frac{\mathrm{h}}{2\pi \mathrm{c%
}}\right) ^3\int_{\left( \mathbf{\Omega }\right) }d\mathbf{x}\cdot e^{%
\mathrm{i}\frac{\mathrm{h}}{\mathrm{c}}\mathbf{kx}}\underline{e}%
_{*r,j}^{\dagger }\left( \mathbf{k}\right) \psi \left( \mathbf{x}\right) 
\mathbf{\mbox{.}}
\end{eqnarray*}

Let

\[
\psi ^{\prime }\left( \mathbf{x}\right) :=U^{\left( -\right) }\psi \left( 
\mathbf{x}\right) \mbox{.} 
\]

In that case:

\begin{eqnarray*}
&&\ \ b_{\circ r,\mathbf{k}}^{\prime }:=\left( \frac{\mathrm{h%
}}{2\pi \mathrm{c}}\right) ^3\int_{\left( \mathbf{\Omega }\right) }d\mathbf{x%
}\cdot e^{\mathrm{i}\frac{\mathrm{h}}{\mathrm{c}}\mathbf{kx}}\underline{e}%
_{\circ r,j}^{\prime \dagger }\left( \mathbf{k}\right) \psi ^{\prime }\left( 
\mathbf{x}\right) \mbox{,} \\
&&\ \ b_{*r,\mathbf{k}}^{\prime }:=\left( \frac{\mathrm{h}}{%
2\pi \mathrm{c}}\right) ^3\int_{\left( \mathbf{\Omega }\right) }d\mathbf{x}%
\cdot e^{\mathrm{i}\frac{\mathrm{h}}{\mathrm{c}}\mathbf{kx}}\underline{e}%
_{*r,j}^{\prime \dagger }\left( \mathbf{k}\right) \psi ^{\prime }\left( 
\mathbf{x}\right) \mathbf{\mbox{.}}
\end{eqnarray*}

Hence:

\begin{eqnarray*}
\ \ \ b_{\circ r,\mathbf{k}}^{\prime } &=&\left( \frac{\mathrm{h}}{2\pi 
\mathrm{c}}\right) ^3\int_{\left( \mathbf{\Omega }\right) }d\mathbf{x}\cdot
e^{\mathrm{i}\frac{\mathrm{h}}{\mathrm{c}}\mathbf{kx}}\left( U^{\left(
-\right) }\underline{e}_{\circ r,j}\left( \mathbf{k}\right) \right)
^{\dagger }\left( U^{\left( -\right) }\psi \left( \mathbf{x}\right) \right) 
\mathbf{\mbox{,}} \\
\ \ \ b_{*r,\mathbf{k}}^{\prime } &=&\left( \frac{\mathrm{h}}{2\pi \mathrm{c}%
}\right) ^3\int_{\left( \mathbf{\Omega }\right) }d\mathbf{x}\cdot e^{\mathrm{%
i}\frac{\mathrm{h}}{\mathrm{c}}\mathbf{kx}}\left( U^{\left( -\right) }%
\underline{e}_{*r,j}\left( \mathbf{k}\right) \right) ^{\dagger }\left(
U^{\left( -\right) }\psi \left( \mathbf{x}\right) \right) \mathbf{\mbox{.}}
\end{eqnarray*}

Since $U^{\left( -\right) \dagger }U^{\left( -\right) }=1_8$ then

\begin{eqnarray*}
\ \ \ \ \ b_{\circ r,\mathbf{k}}^{\prime } &=&\left( \frac{\mathrm{h}}{2\pi 
\mathrm{c}}\right) ^3\int_{\left( \mathbf{\Omega }\right) }d\mathbf{x}\cdot
e^{\mathrm{i}\frac{\mathrm{h}}{\mathrm{c}}\mathbf{kx}}\underline{e}_{\circ
r,j}^{\dagger }\left( \mathbf{k}\right) \psi \left( \mathbf{x}\right) 
\mathbf{\mbox{,}} \\
\ \ \ \ \ b_{*r,\mathbf{k}}^{\prime } &=&\left( \frac{\mathrm{h}}{2\pi 
\mathrm{c}}\right) ^3\int_{\left( \mathbf{\Omega }\right) }d\mathbf{x}\cdot
e^{\mathrm{i}\frac{\mathrm{h}}{\mathrm{c}}\mathbf{kx}}\underline{e}%
_{*r,j}^{\dagger }\left( \mathbf{k}\right) \psi \left( \mathbf{x}\right) 
\mathbf{\mbox{.}}
\end{eqnarray*}

That is:

\[
\ b_{\circ r,\mathbf{k}}^{\prime }=\ b_{\circ r,\mathbf{k}}\mbox{ and }b_{*r,%
\mathbf{k}}^{\prime }=\ b_{*r,\mathbf{k}}\mbox{.} 
\]

And from (\ref{ntr1}):

\begin{equation}
\psi _j^{\prime }\left( \mathbf{x}\right) =\sum_{\mathbf{k}}e^{-\mathrm{i}%
\frac{\mathrm{h}}{\mathrm{c}}\mathbf{kx}}\sum_{r=1}^4\left( \underline{e}%
_{\circ r,j}^{\prime }\left( \mathbf{k}\right) b_{\circ r,\mathbf{k}}+%
\underline{e}_{*r,j}^{\prime }\left( \mathbf{k}\right) b_{*r,\mathbf{k}%
}\right) \mbox{.}  \label{ntr3}
\end{equation}

For the operators\index{operator} $b_{\circ r,\mathbf{k}}$ and $b_{*r,%
\mathbf{k}}$:

\begin{equation}
\frame{$
\begin{array}{c}
\left\{ b_{\circ r^{\prime },\mathbf{{k}^{\prime }}}^{\dagger },b_{\circ r,%
\mathbf{k}}\right\} =\left( \frac{\mathrm{h}}{2\pi \mathrm{c}}\right)
^3\delta _{r,r^{\prime }}\delta _{{\mathbf{k,k^{\prime }}}}\widehat{1}%
\mbox{,} \\ 
\left\{ b_{*r^{\prime },\mathbf{{k}^{\prime }}}^{\dagger },b_{*r,\mathbf{k}%
}\right\} =\left( \frac{\mathrm{h}}{2\pi \mathrm{c}}\right) ^3\delta
_{r,r^{\prime }}\delta _{{\mathbf{k,k^{\prime }}}}\widehat{1}\mbox{,} \\ 
\left\{ b_{\circ r^{\prime },\mathbf{{k}^{\prime }}}^{\dagger },b_{*r,%
\mathbf{k}}\right\} =\widehat{0}\mbox{,} \\ 
\left\{ b_{*r^{\prime },\mathbf{{k}^{\prime }}}^{\dagger },b_{\circ r,%
\mathbf{k}}\right\} =\widehat{0}\mbox{,} \\ 
\left\{ b_{\circ r^{\prime },\mathbf{{k}^{\prime }}}^{\dagger },b_{\circ r,%
\mathbf{k}}^{\dagger }\right\} =\widehat{0}\mbox{,} \\ 
\left\{ b_{*r^{\prime },\mathbf{{k}^{\prime }}}^{\dagger },b_{*r,\mathbf{k}%
}^{\dagger }\right\} =\widehat{0}\mbox{,} \\ 
\left\{ b_{\circ r^{\prime },\mathbf{{k}^{\prime }}},b_{\circ r,\mathbf{k}%
}\right\} =\widehat{0}\mbox{,} \\ 
\left\{ b_{*r^{\prime },\mathbf{{k}^{\prime }}},b_{*r,\mathbf{k}}\right\} =%
\widehat{0}\mbox{,} \\ 
\left\{ b_{*r^{\prime },\mathbf{{k}^{\prime }}}^{\dagger },b_{\circ r,%
\mathbf{k}}^{\dagger }\right\} =\widehat{0}\mbox{.}
\end{array}
$}  \label{box}
\end{equation}

Let

\[
b_{r,\mathbf{k}}:=\sqrt{2}\left( 1-a^2\right) ^{\frac
14}\left( \frac 1{\sqrt{\sqrt{1-a^2}-b}}b_{\circ r,\mathbf{k}}+\frac 1{\sqrt{%
\sqrt{1-a^2}+b}}b_{*r,\mathbf{k}}\right) \mbox{.} 
\]

In that case:

\begin{eqnarray*}
&&\underline{e}_{\circ r}^{\prime }\left( \mathbf{k}\right) b_{\circ r,%
\mathbf{k}}+\underline{e}_{*r}^{\prime }\left( \mathbf{k}\right) b_{*r,%
\mathbf{k}}= \\
&=&\frac 1{\sqrt{2}}\left( \sqrt{1-\frac b{\sqrt{1-a^2}}}\underline{e}%
_{\circ r}^{\prime }\left( \mathbf{k}\right) +\sqrt{1+\frac b{\sqrt{1-a^2}}}%
\underline{e}_{*r}^{\prime }\left( \mathbf{k}\right) \right) b_{r,\mathbf{k}}
\\
&&-\sqrt{\frac{b-\sqrt{1-a^2}}{b+\sqrt{1-a^2}}}\underline{e}_{\circ
r}^{\prime }\left( \mathbf{k}\right) b_{*r,\mathbf{k}}-\sqrt{\frac{b+\sqrt{%
1-a^2}}{b-\sqrt{1-a^2}}}\underline{e}_{*r}^{\prime }\left( \mathbf{k}\right)
b_{\circ r,\mathbf{k}}
\end{eqnarray*}

And from (\ref{ntr2}):

\begin{eqnarray*}
&&\underline{e}_{\circ r}^{\prime }\left( \mathbf{k}\right) b_{\circ r,%
\mathbf{k}}+\underline{e}_{*r}^{\prime }\left( \mathbf{k}\right) b_{*r,%
\mathbf{k}} \\
&=&\underline{e}_r^{\prime }\left( \mathbf{k}\right) b_{r,\mathbf{k}} \\
&&-\sqrt{\frac{b-\sqrt{1-a^2}}{b+\sqrt{1-a^2}}}\underline{e}_{\circ
r}^{\prime }\left( \mathbf{k}\right) b_{*r,\mathbf{k}} \\
&&-\sqrt{\frac{b+\sqrt{1-a^2}}{b-\sqrt{1-a^2}}}\underline{e}_{*r}^{\prime
}\left( \mathbf{k}\right) b_{\circ r,\mathbf{k}}\mbox{.}
\end{eqnarray*}

For $b_{r,\mathbf{k}}$:

\begin{equation}
\frame{$
\begin{array}{c}
\left\{ b_{r^{\prime },\mathbf{k}^{\prime }}^{\dagger },b_{r,\mathbf{k}%
}\right\} =4\frac{b^2+c^2+q^2}{c^2+q^2}\left( \frac h{2\pi \mathrm{c}%
}\right) ^3\delta _{r,r^{\prime }}\delta _{{\mathbf{k,k^{\prime }}}}\widehat{%
1}\mbox{,} \\ 
\left\{ b_{r^{\prime },\mathbf{k}^{\prime }}^{\dagger },b_{r,\mathbf{k}%
}^{\dagger }\right\} =\widehat{0}\mbox{,} \\ 
\left\{ b_{r^{\prime },\mathbf{k}^{\prime }},b_{r,\mathbf{k}}\right\} =%
\widehat{0}\mbox{.}
\end{array}
$}  \label{br}
\end{equation}

From (\ref{ntr3}):

\begin{eqnarray*}
\psi _j^{\prime }\left( \mathbf{x}\right) &=&\sum_{\mathbf{k}}e^{-\mathrm{i}%
\frac{\mathrm{h}}{\mathrm{c}}\mathbf{kx}}\sum_{r=1}^4\underline{e}%
_{r,j}^{\prime }\left( \mathbf{k}\right) b_{r,\mathbf{k}} \\
&&-\sqrt{\frac{b-\sqrt{1-a^2}}{b+\sqrt{1-a^2}}}\sum_{\mathbf{k}}e^{-\mathrm{i%
}\frac{\mathrm{h}}{\mathrm{c}}\mathbf{kx}}\sum_{r=1}^4\underline{e}_{\circ
r,j}^{\prime }\left( \mathbf{k}\right) b_{*r,\mathbf{k}} \\
&&-\sqrt{\frac{b+\sqrt{1-a^2}}{b-\sqrt{1-a^2}}}\sum_{\mathbf{k}}e^{-\mathrm{i%
}\frac{\mathrm{h}}{\mathrm{c}}\mathbf{kx}}\sum_{r=1}^4\underline{e}%
_{*r,j}^{\prime }\left( \mathbf{k}\right) b_{\circ r,\mathbf{k}}\mbox{.}
\end{eqnarray*}

Let:

\begin{equation}
\chi \left( \mathbf{x}\right) :=\sum_{\mathbf{k}}e^{-\mathrm{i%
}\frac{\mathrm{h}}{\mathrm{c}}\mathbf{kx}}\sum_{r=1}^4\underline{e}%
_r^{\prime }\left( \mathbf{k}\right) b_{r,\mathbf{k}}\mbox{,}  \label{ntr5}
\end{equation}

\[
\chi _{*j}\left( \mathbf{x}\right) :=\sqrt{\frac{b-\sqrt{1-a^2%
}}{b+\sqrt{1-a^2}}}\sum_{\mathbf{k}}e^{-\mathrm{i}\frac{\mathrm{h}}{\mathrm{c%
}}\mathbf{kx}}\sum_{r=1}^4\underline{e}_{\circ r,j}^{\prime }\left( \mathbf{k%
}\right) b_{*r,\mathbf{k}} 
\]

\[
\chi _{\circ j}\left( \mathbf{x}\right) :=\sqrt{\frac{b+\sqrt{%
1-a^2}}{b-\sqrt{1-a^2}}}\sum_{\mathbf{k}}e^{-\mathrm{i}\frac{\mathrm{h}}{%
\mathrm{c}}\mathbf{kx}}\sum_{r=1}^4\underline{e}_{*r,j}^{\prime }\left( 
\mathbf{k}\right) b_{\circ r,\mathbf{k}}\mbox{.} 
\]

In that case:

\[
\psi _j^{\prime }\left( \mathbf{x}\right) =\chi _j\left( \mathbf{x}\right)
-\chi _{*j}\left( \mathbf{x}\right) -\chi _{\circ j}\left( \mathbf{x}\right)%
\mbox{.} 
\]

Let

\[
\widehat{H}_{0}^{\prime }:=U^{\left( -\right) }\widehat{H}%
_{0}U^{\left( -\right) \dagger } \mbox{.} 
\]

For this Hamiltonian\index{Hamiltonian}:

\begin{eqnarray*}
&&\int_{\left( \mathbf{\Omega }\right) }d\mathbf{x}\cdot \chi _{*}^{\dagger
}\left( \mathbf{x}\right) \widehat{H}_0^{\prime }\psi ^{\prime }\left( 
\mathbf{x}\right) \\
&=&\int_{\left( \mathbf{\Omega }\right) }d\mathbf{x}\cdot \sqrt{\frac{b-%
\sqrt{1-a^2}}{b+\sqrt{1-a^2}}}\sum_{\mathbf{k}^{\prime }}e^{\mathrm{i}\frac{%
\mathrm{h}}{\mathrm{c}}\mathbf{k}^{\prime }\mathbf{x}}\sum_{r^{\prime
}=1}^4b_{*r^{\prime },\mathbf{k}^{\prime }}^{\dagger }\underline{e}_{\circ
r^{\prime }}^{\prime \dagger }\left( \mathbf{k}^{\prime }\right) \cdot \\
&&\cdot \widehat{H}_0^{\prime }\sum_{\mathbf{k}}e^{-\mathrm{i}\frac{\mathrm{h%
}}{\mathrm{c}}\mathbf{kx}}\sum_{r=1}^4\left( \underline{e}_{\circ r}^{\prime
}\left( \mathbf{k}\right) b_{\circ r,\mathbf{k}}+\underline{e}_{*r}^{\prime
}\left( \mathbf{k}\right) b_{*r,\mathbf{k}}\right) \\
&=&\int_{\left( \mathbf{\Omega }\right) }d\mathbf{x}\cdot \sqrt{\frac{b-%
\sqrt{1-a^2}}{b+\sqrt{1-a^2}}}\sum_{\mathbf{k}}\sum_{\mathbf{k}^{\prime }}e^{%
\mathrm{i}\frac{\mathrm{h}}{\mathrm{c}}\mathbf{k}^{\prime }\mathbf{x}%
}\sum_{r^{\prime }=1}^4b_{*r^{\prime },\mathbf{k}^{\prime }}^{\dagger }%
\underline{e}_{\circ r^{\prime }}^{\prime \dagger }\left( \mathbf{k}^{\prime
}\right) \cdot \\
&&\cdot \widehat{H}_0^{\prime }e^{-\mathrm{i}\frac{\mathrm{h}}{\mathrm{c}}%
\mathbf{kx}}\left( 
\begin{array}{c}
\sum_{r=1}^2\left( \underline{e}_{\circ r}^{\prime }\left( \mathbf{k}\right)
b_{\circ r,\mathbf{k}}+\underline{e}_{*r}^{\prime }\left( \mathbf{k}\right)
b_{*r,\mathbf{k}}\right) + \\ 
+\sum_{r=3}^4\left( \underline{e}_{\circ r}^{\prime }\left( \mathbf{k}%
\right) b_{\circ r,\mathbf{k}}+\underline{e}_{*r}^{\prime }\left( \mathbf{k}%
\right) b_{*r,\mathbf{k}}\right)
\end{array}
\right) \\
&=&\int_{\left( \mathbf{\Omega }\right) }d\mathbf{x}\cdot \sqrt{\frac{b-%
\sqrt{1-a^2}}{b+\sqrt{1-a^2}}}\sum_{\mathbf{k}}\sum_{\mathbf{k}^{\prime }}e^{%
\mathrm{i}\frac{\mathrm{h}}{\mathrm{c}}\mathbf{k}^{\prime }\mathbf{x}}e^{-%
\mathrm{i}\frac{\mathrm{h}}{\mathrm{c}}\mathbf{kx}}\sum_{r^{\prime
}=1}^4b_{*r^{\prime },\mathbf{k}^{\prime }}^{\dagger }\underline{e}_{\circ
r^{\prime }}^{\prime \dagger }\left( \mathbf{k}^{\prime }\right) \cdot \\
&&\cdot \mathrm{h}\widehat{H}_0^{\prime }\left( \mathbf{k}\right) \left( 
\begin{array}{c}
\sum_{r=1}^2\left( \underline{e}_{\circ r}^{\prime }\left( \mathbf{k}\right)
b_{\circ r,\mathbf{k}}+\underline{e}_{*r}^{\prime }\left( \mathbf{k}\right)
b_{*r,\mathbf{k}}\right) + \\ 
+\sum_{r=3}^4\left( \underline{e}_{\circ r}^{\prime }\left( \mathbf{k}%
\right) b_{\circ r,\mathbf{k}}+\underline{e}_{*r}^{\prime }\left( \mathbf{k}%
\right) b_{*r,\mathbf{k}}\right)
\end{array}
\right) \mbox{.}
\end{eqnarray*}

Hence:

\begin{eqnarray*}
&&\int_{\left( \mathbf{\Omega }\right) }d\mathbf{x}\cdot \chi _{*}^{\dagger
}\left( \mathbf{x}\right) \widehat{H}_0^{\prime }\psi ^{\prime }\left( 
\mathbf{x}\right) \\
&=&\int_{\left( \mathbf{\Omega }\right) }d\mathbf{x}\cdot \sqrt{\frac{b-%
\sqrt{1-a^2}}{b+\sqrt{1-a^2}}}\sum_{\mathbf{k}}\sum_{\mathbf{k}^{\prime }}e^{%
\mathrm{i}\frac{\mathrm{h}}{\mathrm{c}}\mathbf{k}^{\prime }\mathbf{x}}e^{-%
\mathrm{i}\frac{\mathrm{h}}{\mathrm{c}}\mathbf{kx}}\sum_{r^{\prime
}=1}^4b_{*r^{\prime },\mathbf{k}^{\prime }}^{\dagger }\underline{e}_{\circ
r^{\prime }}^{\prime \dagger }\left( \mathbf{k}^{\prime }\right) \cdot \\
&&\cdot \mathrm{h}\left( 
\begin{array}{c}
\omega \left( \mathbf{k}\right) \sum_{r=1}^2\left( \underline{e}_{\circ
r}^{\prime }\left( \mathbf{k}\right) b_{\circ r,\mathbf{k}}+\underline{e}%
_{*r}^{\prime }\left( \mathbf{k}\right) b_{*r,\mathbf{k}}\right) - \\ 
-\omega \left( \mathbf{k}\right) \sum_{r=3}^4\left( \underline{e}_{\circ
r}^{\prime }\left( \mathbf{k}\right) b_{\circ r,\mathbf{k}}+\underline{e}%
_{*r}^{\prime }\left( \mathbf{k}\right) b_{*r,\mathbf{k}}\right)
\end{array}
\right) \\
&=&\int_{\left( \mathbf{\Omega }\right) }d\mathbf{x}\cdot \sqrt{\frac{b-%
\sqrt{1-a^2}}{b+\sqrt{1-a^2}}}\sum_{\mathbf{k}}\mathrm{h}\omega \left( 
\mathbf{k}\right) \sum_{\mathbf{k}^{\prime }}e^{\mathrm{i}\frac{\mathrm{h}}{%
\mathrm{c}}\mathbf{k}^{\prime }\mathbf{x}}e^{-\mathrm{i}\frac{\mathrm{h}}{%
\mathrm{c}}\mathbf{kx}}\cdot \\
&&\cdot \sum_{r^{\prime }=1}^4b_{*r^{\prime },\mathbf{k}^{\prime }}^{\dagger
}\underline{e}_{\circ r^{\prime }}^{\prime \dagger }\left( \mathbf{k}%
^{\prime }\right) \left( 
\begin{array}{c}
\sum_{r=1}^2\left( \underline{e}_{\circ r}^{\prime }\left( \mathbf{k}\right)
b_{\circ r,\mathbf{k}}+\underline{e}_{*r}^{\prime }\left( \mathbf{k}\right)
b_{*r,\mathbf{k}}\right) - \\ 
-\sum_{r=3}^4\left( \underline{e}_{\circ r}^{\prime }\left( \mathbf{k}%
\right) b_{\circ r,\mathbf{k}}+\underline{e}_{*r}^{\prime }\left( \mathbf{k}%
\right) b_{*r,\mathbf{k}}\right)
\end{array}
\right) \\
&=&\sqrt{\frac{b-\sqrt{1-a^2}}{b+\sqrt{1-a^2}}}\sum_{\mathbf{k}}\mathrm{h}%
\omega \left( \mathbf{k}\right) \sum_{\mathbf{k}^{\prime }}\left( \int d%
\mathbf{x}\cdot e^{-\mathrm{i}\frac{\mathrm{h}}{\mathrm{c}}\left( \mathbf{k-k%
}^{\prime }\right) \mathbf{x}}\right) \cdot \\
&&\cdot \sum_{r^{\prime }=1}^4b_{*r^{\prime },\mathbf{k}^{\prime }}^{\dagger
}\left( 
\begin{array}{c}
\sum_{r=1}^2\left( \underline{e}_{\circ r^{\prime }}^{\prime \dagger }\left( 
\mathbf{k}^{\prime }\right) \underline{e}_{\circ r}^{\prime }\left( \mathbf{k%
}\right) b_{\circ r,\mathbf{k}}+\underline{e}_{\circ r^{\prime }}^{\prime
\dagger }\left( \mathbf{k}^{\prime }\right) \underline{e}_{*r}^{\prime
}\left( \mathbf{k}\right) b_{*r,\mathbf{k}}\right) - \\ 
-\sum_{r=3}^4\left( \underline{e}_{\circ r^{\prime }}^{\prime \dagger
}\left( \mathbf{k}^{\prime }\right) \underline{e}_{\circ r}^{\prime }\left( 
\mathbf{k}\right) b_{\circ r,\mathbf{k}}+\underline{e}_{\circ r^{\prime
}}^{\prime \dagger }\left( \mathbf{k}^{\prime }\right) \underline{e}%
_{*r}^{\prime }\left( \mathbf{k}\right) b_{*r,\mathbf{k}}\right)
\end{array}
\right) \mbox{.}
\end{eqnarray*}

Since

\[
\int_{\left( \mathbf{\Omega }\right) }d\mathbf{x}\cdot e^{-\mathrm{i}\frac{%
\mathrm{h}}{\mathrm{c}}\left( \mathbf{k-k}^{\prime }\right) \mathbf{x}%
}=\left( \frac{2\pi \mathrm{c}}{\mathrm{h}}\right) ^3\delta _{\mathbf{k,k}%
^{\prime }} 
\]

then

\begin{eqnarray*}
&&\int_{\left( \mathbf{\Omega }\right) }d\mathbf{x}\cdot \chi _{*}^{\dagger
}\left( \mathbf{x}\right) \widehat{H}_0^{\prime }\psi ^{\prime }\left( 
\mathbf{x}\right) \\
&=&\sqrt{\frac{b-\sqrt{1-a^2}}{b+\sqrt{1-a^2}}}\sum_{\mathbf{k}}\mathrm{h}%
\omega \left( \mathbf{k}\right) \sum_{\mathbf{k}^{\prime }}\left( \frac{2\pi 
\mathrm{c}}{\mathrm{h}}\right) ^3\delta _{\mathbf{k,k}^{\prime }}\cdot \\
&&\cdot \sum_{r^{\prime }=1}^4b_{*r^{\prime },\mathbf{k}^{\prime }}^{\dagger
}\left( 
\begin{array}{c}
\sum_{r=1}^2\left( \underline{e}_{\circ r^{\prime }}^{\prime \dagger }\left( 
\mathbf{k}^{\prime }\right) \underline{e}_{\circ r}^{\prime }\left( \mathbf{k%
}\right) b_{\circ r,\mathbf{k}}+\underline{e}_{\circ r^{\prime }}^{\prime
\dagger }\left( \mathbf{k}^{\prime }\right) \underline{e}_{*r}^{\prime
}\left( \mathbf{k}\right) b_{*r,\mathbf{k}}\right) - \\ 
-\sum_{r=3}^4\left( \underline{e}_{\circ r^{\prime }}^{\prime \dagger
}\left( \mathbf{k}^{\prime }\right) \underline{e}_{\circ r}^{\prime }\left( 
\mathbf{k}\right) b_{\circ r,\mathbf{k}}+\underline{e}_{\circ r^{\prime
}}^{\prime \dagger }\left( \mathbf{k}^{\prime }\right) \underline{e}%
_{*r}^{\prime }\left( \mathbf{k}\right) b_{*r,\mathbf{k}}\right)
\end{array}
\right) \mbox{.}
\end{eqnarray*}

In accordance with properties of $\delta $:

\begin{eqnarray*}
&&\int_{\left( \mathbf{\Omega }\right) }d\mathbf{x}\cdot \chi _{*}^{\dagger
}\left( \mathbf{x}\right) \widehat{H}_0^{\prime }\psi ^{\prime }\left( 
\mathbf{x}\right) \\
&=&\sqrt{\frac{b-\sqrt{1-a^2}}{b+\sqrt{1-a^2}}}\sum_{\mathbf{k}}\mathrm{h}%
\omega \left( \mathbf{k}\right) \left( \frac{2\pi \mathrm{c}}{\mathrm{h}}%
\right) ^3\cdot \\
&&\cdot \sum_{r^{\prime }=1}^4b_{*r^{\prime },\mathbf{k}}^{\dagger }\left( 
\begin{array}{c}
\sum_{r=1}^2\left( \underline{e}_{\circ r^{\prime }}^{\prime \dagger }\left( 
\mathbf{k}\right) \underline{e}_{\circ r}^{\prime }\left( \mathbf{k}\right)
b_{\circ r,\mathbf{k}}+\underline{e}_{\circ r^{\prime }}^{\prime \dagger
}\left( \mathbf{k}\right) \underline{e}_{*r}^{\prime }\left( \mathbf{k}%
\right) b_{*r,\mathbf{k}}\right) - \\ 
-\sum_{r=3}^4\left( \underline{e}_{\circ r^{\prime }}^{\prime \dagger
}\left( \mathbf{k}\right) \underline{e}_{\circ r}^{\prime }\left( \mathbf{k}%
\right) b_{\circ r,\mathbf{k}}+\underline{e}_{\circ r^{\prime }}^{\prime
\dagger }\left( \mathbf{k}\right) \underline{e}_{*r}^{\prime }\left( \mathbf{%
k}\right) b_{*r,\mathbf{k}}\right)
\end{array}
\right) \mbox{.}
\end{eqnarray*}

Since

\begin{eqnarray*}
\underline{e}_{\circ r^{\prime }}^{\prime \dagger }\left( \mathbf{k}^{\prime
}\right) \underline{e}_{\circ r}^{\prime }\left( \mathbf{k}\right) &=&\delta
_{r,r^{\prime }}\mbox{,} \\
\underline{e}_{\circ r^{\prime }}^{\prime \dagger }\left( \mathbf{k}^{\prime
}\right) \underline{e}_{*r}^{\prime }\left( \mathbf{k}\right) &=&0
\end{eqnarray*}

then

\begin{eqnarray*}
&&\int_{\left( \mathbf{\Omega }\right) }d\mathbf{x}\cdot \chi _{*}^{\dagger
}\left( \mathbf{x}\right) \widehat{H}_0^{\prime }\psi ^{\prime }\left( 
\mathbf{x}\right) \\
&=&\sqrt{\frac{b-\sqrt{1-a^2}}{b+\sqrt{1-a^2}}}\sum_{\mathbf{k}}\mathrm{h}%
\omega \left( \mathbf{k}\right) \left( \frac{2\pi \mathrm{c}}{\mathrm{h}}%
\right) ^3\cdot \\
&&\cdot \sum_{r^{\prime }=1}^4b_{*r^{\prime },\mathbf{k}}^{\dagger }\left(
\sum_{r=1}^2\left( \delta _{r,r^{\prime }}b_{\circ r,\mathbf{k}}+0b_{*r,%
\mathbf{k}}\right) -\sum_{r=3}^4\left( \delta _{r,r^{\prime }}b_{\circ r,%
\mathbf{k}}+0b_{*r,\mathbf{k}}\right) \right) \\
&=&\left( \frac{2\pi \mathrm{c}}{\mathrm{h}}\right) ^3\sqrt{\frac{b-\sqrt{%
1-a^2}}{b+\sqrt{1-a^2}}}\sum_{\mathbf{k}}\mathrm{h}\omega \left( \mathbf{k}%
\right) \cdot \\
&&\cdot \sum_{r^{\prime }=1}^4b_{*r^{\prime },\mathbf{k}}^{\dagger }\left(
\sum_{r=1}^2\delta _{r,r^{\prime }}b_{\circ r,\mathbf{k}}-\sum_{r=3}^4\delta
_{r,r^{\prime }}b_{\circ r,\mathbf{k}}\right) \\
&=&\left( \frac{2\pi \mathrm{c}}{\mathrm{h}}\right) ^3\sqrt{\frac{b-\sqrt{%
1-a^2}}{b+\sqrt{1-a^2}}}\cdot \\
&&\cdot \sum_{\mathbf{k}}\mathrm{h}\omega \left( \mathbf{k}\right) \left(
\sum_{r=1}^2b_{*r,\mathbf{k}}^{\dagger }b_{\circ r,\mathbf{k}%
}-\sum_{r=3}^4b_{*r,\mathbf{k}}^{\dagger }b_{\circ r,\mathbf{k}}\right) %
\mbox{.}
\end{eqnarray*}

Therefore

\begin{eqnarray*}
&&\int_{\left( \mathbf{\Omega }\right) }d\mathbf{x}\cdot \chi _{*}^{\dagger
}\left( \mathbf{x}\right) \widehat{H}_0^{\prime }\psi ^{\prime }\left( 
\mathbf{x}\right) \\
&=&\left( \frac{2\pi \mathrm{c}}{\mathrm{h}}\right) ^3\sqrt{\frac{b-\sqrt{%
1-a^2}}{b+\sqrt{1-a^2}}}\sum_{\mathbf{k}}\mathrm{h}\omega \left( \mathbf{k}%
\right) \left( \sum_{r=1}^2b_{*r,\mathbf{k}}^{\dagger }b_{\circ r,\mathbf{k}%
}-\sum_{r=3}^4b_{*r,\mathbf{k}}^{\dagger }b_{\circ r,\mathbf{k}}\right) %
\mbox{.}
\end{eqnarray*}

Similarly you can calculate that

\begin{eqnarray*}
&&\ \int d\mathbf{x}\cdot \chi _{\circ }^{\dagger }\left( \mathbf{x}\right) 
\widehat{H}_0^{\prime }\psi ^{\prime }\left( \mathbf{x}\right) \\
\ &=&\left( \frac{2\pi \mathrm{c}}{\mathrm{h}}\right) ^3\sqrt{\frac{b-\sqrt{%
1-a^2}}{b+\sqrt{1-a^2}}}\sum_{\mathbf{k}}\mathrm{h}\omega \left( \mathbf{k}%
\right) \left( \sum_{r=1}^2b_{\circ r,\mathbf{k}}^{\dagger }b_{*r,\mathbf{k}%
}-\sum_{r=3}^4b_{\circ r,\mathbf{k}}^{\dagger }b_{*r,\mathbf{k}}\right) %
\mbox{.}
\end{eqnarray*}

Since \cite{Q94}

\[
\widetilde{\Psi }\left( t,\mathbf{p}\right) =\left( \frac{2\pi \mathrm{c}}{%
\mathrm{h}}\right) ^3\sum_{r=1}^4c_r\left( t,\mathbf{p}\right) b_{r,\mathbf{p%
}}^{\dagger }{\widetilde{F}}_0 
\]

and (\ref{box})

\[
\left\{ b_{*r^{\prime },\mathbf{{k}^{\prime }}}^{\dagger },b_{\circ r,%
\mathbf{k}}\right\} =\widehat{0}\mbox{, }\left\{ b_{*r^{\prime },\mathbf{{k}%
^{\prime }}}^{\dagger },b_{*r,\mathbf{k}}^{\dagger }\right\} =\widehat{0}%
\mbox{, }\left\{ b_{*r^{\prime },\mathbf{{k}^{\prime }}}^{\dagger },b_{\circ
r,\mathbf{k}}^{\dagger }\right\} =\widehat{0}\mbox{. } 
\]

then

\[
b_{*r,\mathbf{k}}^{\dagger }b_{\circ r,\mathbf{k}}\widetilde{\Psi }%
=-b_{\circ r,\mathbf{k}}\left( \frac{2\pi \mathrm{c}}{\mathrm{h}}\right)
^3\sum_{r=1}^4c_r\left( t,\mathbf{p}\right) b_{*r,\mathbf{k}}^{\dagger }b_{r,%
\mathbf{p}}^{\dagger }{\widetilde{F}}_0=\widetilde{0}\mbox{.} 
\]

Similarly

\[
b_{\circ r,\mathbf{k}}^{\dagger }b_{*r,\mathbf{k}}\widetilde{\Psi }=%
\widetilde{0}. 
\]

Hence

\[
\frame{$\int_{\left( \mathbf{\Omega }\right) }d\mathbf{x}\cdot \psi ^{\prime
\dagger }\left( \mathbf{x}\right) \widehat{H}_0^{\prime }\psi^{\prime}
\left( \mathbf{x}\right) \Psi \left( t,\mathbf{x}_0\right) =\int_{\left( 
\mathbf{\Omega }\right) }d\mathbf{x\cdot }\chi ^{\dagger }\left( \mathbf{x}%
\right) \widehat{H}_0^{\prime }\chi\left( \mathbf{x}\right) \Psi \left( t,%
\mathbf{x}_0\right) $.} 
\]

Thus, the function $\psi ^{\prime }\left( \mathbf{x}\right) $ can be
substituted for the function $\chi\left( \mathbf{x}\right) $ in calculations
of a probabilities evolution.

Let

\[
\nu _{n_0,\left( s\right) }\left( \mathbf{k}\right) :=\left[ 
\begin{array}{c}
\left( c+iq\right) e_{sL}\left( \mathbf{k}\right) \\ 
\overrightarrow{0}_2
\end{array}
\right] \mbox{, }l_{n_0,\left( s\right) }\left( \mathbf{k}\right):=\left[ 
\begin{array}{c}
\left( a-ib\right) e_{sL}\left( \mathbf{k}\right) \\ 
e_{sR}\left( \mathbf{k}\right)
\end{array}
\right] \mbox{.} 
\]

Hence from (\ref{ntr2}):

\[
\begin{array}{c}
\underline{e}_s^{\prime }\left( \mathbf{k}\right) =\frac{\mathrm{h}}{2\pi 
\mathrm{c}}\cdot \\ 
\cdot \left( 
\begin{array}{c}
\sqrt{\frac{2\pi n_0}{\sinh \left( 2n_0\pi \right) }}\left( 
\begin{array}{c}
\left( 
\begin{array}{c}
\cosh \left( \frac{\mathrm{h}}{\mathrm{c}}n_0x_4\right) + \\ 
+\sinh \left( \frac{\mathrm{h}}{\mathrm{c}}n_0x_4\right)
\end{array}
\right) \left( c+iq\right) e_{sL}\left( \mathbf{k}\right) + \\ 
+\left( 
\begin{array}{c}
\cosh \left( \frac{\mathrm{h}}{\mathrm{c}}n_0x_4\right) - \\ 
-\sinh \left( \frac{\mathrm{h}}{\mathrm{c}}n_0x_4\right)
\end{array}
\right) \overrightarrow{0}_2
\end{array}
\right) + \\ 
+\exp \left( -\mathrm{i}\frac{\mathrm{h}}{\mathrm{c}}\left( n_0x_4\right)
\right) l_{n_0,\left( s\right) }\left( \mathbf{k}\right)
\end{array}
\right) \mbox{.}
\end{array}
\]

Therefore in the basis\index{basis} $\mathbf{J}_{e\nu }$:

\[
\underline{e}_s^{\prime }\left( \mathbf{k}\right) =\left[ 
\begin{array}{c}
\nu _{n_0,\left( s\right) }\left( \mathbf{k}\right) \\ 
l_{n_0,\left( s\right) }\left( \mathbf{k}\right)
\end{array}
\right] \mbox{.} 
\]

Therefore from (\ref{ntr5}):

\[
\chi \left( \mathbf{x}\right) =\sum_{\mathbf{k}}e^{-\mathrm{i}\frac{\mathrm{h%
}}{\mathrm{c}}\mathbf{kx}}\sum_{s=1}^4\left[ 
\begin{array}{c}
\nu _{n_0,\left( s\right) }\left( \mathbf{k}\right) \\ 
l_{n_0,\left( s\right) }\left( \mathbf{k}\right)
\end{array}
\right] b_{s,\mathbf{k}}\mbox{,} 
\]

Let

\[
\widetilde{\nu }_{n_0}\left( \mathbf{x}\right) :=\sum_{%
\mathbf{k}}e^{-\mathrm{i}\frac{\mathrm{h}}{\mathrm{c}}\mathbf{kx}%
}\sum_{s=1}^2\nu _{n_0,\left( s\right) }\left( \mathbf{k}\right) b_s\left( 
\mathbf{k}\right) \mbox{,} 
\]

\[
\widetilde{l}_{n_0}\left( \mathbf{x}\right) :=\sum_{\mathbf{k}%
}e^{-\mathrm{i}\frac{\mathrm{h}}{\mathrm{c}}\mathbf{kx}}\sum_{s=1}^4l_{n_0,%
\left( s\right) }\left( \mathbf{k}\right) b_s\left( \mathbf{k}\right) %
\mbox{.} 
\]

Hence in the basis\index{basis} $\mathbf{J}_{e\nu }$:

\begin{equation}
\chi \left( \mathbf{x}\right) =\left[ 
\begin{array}{c}
\widetilde{\nu }_{n_0}\left( \mathbf{x}\right) \\ 
\widetilde{l}_{n_0}\left( \mathbf{x}\right)
\end{array}
\right] \mbox{.}  \label{Xi}
\end{equation}

Let:

\[
\widehat{H}_{l,0}:=\mathrm{c}\sum_{r=1}^3\beta ^{\left[
r\right] }\mathrm{i}\partial _r+\mathrm{h}n_0\left( a\gamma ^{\left[
0\right] }-b\beta ^{\left[ 4\right] }\right) \mbox{,} 
\]

\[
\widehat{H}_{\nu ,0}:=\mathrm{c}\sum_{r=1}^3\beta ^{\left[
r\right] }\mathrm{i}\partial _r+\mathrm{h}n_0\left( a\gamma ^{\left[
0\right] }+b\beta ^{\left[ 4\right] }\right) \mbox{,} 
\]

\[
\widehat{H}_{\nu ,l}:=\left( c+\mathrm{i}q\right) \left[ 
\begin{array}{cccc}
0 & 0 & n_0 & 0 \\ 
0 & 0 & 0 & n_0 \\ 
-n_0 & 0 & 0 & 0 \\ 
0 & -n_0 & 0 & 0
\end{array}
\right] \mbox{,} 
\]

\[
\widehat{H}_{l,\nu }:=\left( c-\mathrm{i}q\right) \left[ 
\begin{array}{cccc}
0 & 0 & -n_0 & 0 \\ 
0 & 0 & 0 & -n_0 \\ 
n_0 & 0 & 0 & 0 \\ 
0 & n_0 & 0 & 0
\end{array}
\right] \mbox{.} 
\]

In that case in the basis\index{basis} $\mathbf{J}_{e\nu }$:

\[
\widehat{H}_0^{\prime }=\left[ 
\begin{array}{cc}
\widehat{H}_{\nu ,0} & \widehat{H}_{\nu ,l} \\ 
\widehat{H}_{l,\nu } & \widehat{H}_{l,0}
\end{array}
\right] \mbox{.} 
\]

Let

\[
\widehat{H}_{l,0}\left( \mathbf{k}\right) :=\sum_{r=1}^3\beta
^{\left[ r\right] }k_r+n_0\left( a\gamma ^{\left[ 0\right] }-b\beta ^{\left[
4\right] }\right) \mbox{,} 
\]

\[
\widehat{H}_{\nu ,0}\left( \mathbf{k}\right) :=%
\sum_{r=1}^3\beta ^{\left[ r\right] }k_r+n_0\left( a\gamma ^{\left[ 0\right]
}+b\beta ^{\left[ 4\right] }\right) \mbox{.} 
\]

In that case

\[
\widehat{H}_0^{\prime }\left( \mathbf{k}\right) =\left[ 
\begin{array}{cc}
\widehat{H}_{\nu ,0}\left( \mathbf{k}\right) & \widehat{H}_{\nu ,l} \\ 
\widehat{H}_{l,\nu } & \widehat{H}_{l,0}\left( \mathbf{k}\right)
\end{array}
\right] 
\]

An neutrino\index{neutrino} and it's lepton\index{lepton} are tied by the
follows equations:

\[
\widehat{H}_{\nu ,0}\left( \mathbf{k}\right) \nu _{n_0,\left( s\right)
}\left( \mathbf{k}\right) +\widehat{H}_{\nu ,l}l_{n_0,\left( s\right)
}\left( \mathbf{k}\right) =\omega \left( \mathbf{k}\right) \nu _{n_0,\left(
s\right) }\left( \mathbf{k}\right) 
\]

for $s\in \left\{ 1,2\right\} $ and

\[
\widehat{H}_{\nu ,0}\left( \mathbf{k}\right) \nu _{n_0,\left( s\right)
}\left( \mathbf{k}\right) +\widehat{H}_{\nu ,l}l_{n_0,\left( s\right)
}\left( \mathbf{k}\right) =-\omega \left( \mathbf{k}\right) \nu _{n_0,\left(
s\right) }\left( \mathbf{k}\right) 
\]

for $s\in \left\{ 3,4\right\} $.

I suppose that such neutrino can fly 1.5 cm. \cite{CDF} and give birth to it's 
leptons.


\begin{thebibliography}{9}

\bibitem{Q1} G. Quznetsov, {\it Probabilistic Treatment of Gauge Theories}, 
in series Contemporary Fundamental Physics, ed. V. Dvoeglazov, Nova Sci. Publ., 
NY (2007), p. 98

\bibitem{Q2} Idem, p.p. 72, 75, 76 

\bibitem{Q3} Idem, p.p. 98, 99

\bibitem{Q4} Idem, p. 77

\bibitem{Q94} Idem, p. 94

\bibitem{CDF} CDF Collaboration, Study of multi-muon events 
produced in pôp collisions at ps = 1.96 TeV, {\it preprint} \\
http://arxiv.org/abs/0810.5357

\end{thebibliography}
\end{document}